\documentclass[aps,prc,superscriptaddress,showpacs,floatfix,notitlepage,twocolumn,nofootinbib,preprintnumbers]{revtex4-2}
\usepackage{amsmath,graphicx,float,csquotes,mdframed,appendix}

\usepackage[colorlinks=true,linkcolor=blue,citecolor=blue, urlcolor=blue]{hyperref}
\usepackage{cleveref}
\usepackage{amssymb}
\usepackage{pifont}

\newcommand{\nene}{$^{20}$Ne$^{20}$Ne}
\newcommand{\oooo}{$^{16}$O$^{16}$O}

\newcommand{\oo}{$^{16}$O}

\newcommand{\nee}{$^{20}$Ne}
\newcommand{\pb}{$^{208}$Pb}
\newcommand{\xe}{$^{129}$Xe}

\newcommand{\jewel}{JEWEL\mbox{}}

\begin{document}
 
%\title{Three models for charged hadron nuclear modification from light to heavy ions}
%\title{Three predictions for charged hadron nuclear modification from light to heavy ions}
% \title{System-Size Dependence of Partonic Energy Loss from Three Models}
\title{Three Predictions for Partonic Energy Loss from Light to Heavy Ion Collisions}

\author{Wilke van der Schee}
\affiliation{Theoretical Physics Department, CERN, CH-1211 Gen\`eve 23, Switzerland}
\affiliation{Institute for Theoretical Physics, Utrecht University, 3584 CC Utrecht, The Netherlands}
\affiliation{NIKHEF, Science Park 105, 1098 XG Amsterdam, The Netherlands}
\author{Isobel Kolb\'e}
\affiliation{School of Physics, University of the Witwatersrand, 1 Jan Smuts Ave, Braamfontein, 2000, South-Africa}
\affiliation{Mandelstam Institute for Theoretical Physics, University of the Witwatersrand, 1 Jan Smuts Ave, Braamfontein, 2000, South-Africa}
\affiliation{National Institute for Theoretical and Computational Sciences,Merensky building, Merriman Street, Stellenbosch, 7600}
\author{Govert Nijs}
\affiliation{Theoretical Physics Department, CERN, CH-1211 Gen\`eve 23, Switzerland}
\author{Kumail  Ruhani}
\affiliation{Department of Physics, Quaid-i-Azam University, Islamabad}
\author{Ishtiaq Ahmed}
\affiliation{National Centre for Physics, Shahdra Valley Road, Islamabad}
\author{Shahin Iqbal}
\affiliation{National Centre for Physics, Shahdra Valley Road, Islamabad}

\begin{abstract}

\noindent In July 2025 the Large Hadron Collider (LHC) collided $^{16}$O$^{16}$O and $^{20}$Ne$^{20}$Ne isotopes in a quest to understand the physics of ultrarelativistic light ion collisions. 
One of the key motivations for this run is to discover partonic energy loss in systems with a small quark-gluon plasma (QGP). 
In this letter we combine a BDMPS-Z based model, a weighted path based energy loss prescription, and \jewel{}
together with two realistic geometries of the \oo{} and \nee{} isotopes. 
The different sizes of the ions affect the energy loss in characteristically different ways depending on the model.
\end{abstract}

\preprint{CERN-TH-2025-182}

\maketitle

\section{Introduction\label{sec:intorduction}}

    Ultrarelativistic light-ion collisions at the Large Hadron Collider (LHC) are bringing the collider into a new experimental phase that will not only enhance the discovery potential of the heavy-ion programme 
    \cite{Nijs:2025qxm,vanderSchee:2023uii}, but broadens the impact of particle and collider physics as a whole.
    In this context there has been an effort recently from the nuclear structure community    
    to  understand the shapes of the isotopes \oo{} and \nee{} precisely, which is both interesting and necessary to understand the upcoming collisions \cite{Giacalone:2024luz,Giacalone:2024ixe,Summerfield:2021oex,YuanyuanWang:2024sgp}.
    The collisions will have an impact on cosmic-ray physics 
    \cite[Section 11.3]{Citron:2018lsq} 
    %\cite{Citron:2018lsq} 
    as well as synergy with the same collisions  performed at RHIC at a lower collision energy \cite{Brewer:2021kiv,Huang:2023viw}.
        
    The light-ion run will enhance our understanding of relativistic hydrodynamics at its limits (as demonstrated by the first low-$p_T$ results from \oooo{} runs at RHIC and the LHC that appeared after the first release of this letter \cite{ATLAS:2025nnt,ALICE:2025luc,CMS:2025tga,STAR:2025ivi}), as well as jet-medium interactions in small droplets of    quark-gluon plasma (QGP) \cite{Grosse-Oetringhaus:2024bwr}. 
    A particular hope for the light-ion run at the LHC is that it will shed light on the question: ``What is the smallest droplet of deconfined quark matter?'' 
    The question lingers, since in (high multiplicity) proton-proton and proton-lead collisions several telltale signs such as collectivity,
    strangeness enhancement and $J/\psi$-melting have been seen
    \cite{CMS:2010ifv,ALICE:2016fzo,ALICE:2013snh,ALICE:2014cgk, ALICE:2020msa} 
    and studied \cite{Weller:2017tsr, Zhao:2017rgg, Bierlich:2018xfw}, but it has not been possible to observe in-medium parton energy loss \cite{Huss:2020dwe}.
    Crucially, non-zero energy loss is expected in such small systems from a variety of theoretical approaches that successfully describe energy loss in large systems \cite{Faraday:2024qtl}.
    
    Part of the difficulty with observing the modification of hard partons due to the presence of the medium, often called jet-quenching, is that any quenching must necessarily be small, since the quenching is expected to depend on the distance travelled by a hard parton through the medium.
    The measurement of such a small signal is hampered by the need to select events with large event-activity, which correlates with the production of such partons and hence results in large systematic uncertainties \cite{ALICE:2016faw,ALICE:2017svf,ALICE:2014xsp}.
    
    Excitingly, many of these experimental difficulties can be avoided entirely in minimum-bias light-ion collisions.
    A further theoretical and experimental reduction of the uncertainties is possible when taking ratios of observables in different but similar colliding systems, such as will now be possible with \oooo\@ and \nene\@\cite{Giacalone:2024luz,ATLAS:2025nnt,ALICE:2025luc,CMS:2025tga}.
    Freed from the experimental burden of attempting to select relevant events in which a medium with small initial spatial extent is created, the theoretical modelling of jet-quenching in small systems is more easily aligned with experimental measurements.
 
    The charged hadron nuclear modification factor has long been the gold-standard observable for jet quenching, enjoying success in a variety of large collision systems, and highlighting precisely the absence of such in small systems \cite{ATLAS:2022kqu,CMS:2016xef,ALICE:2018vuu,ALICE:2014nqx}.
    In this letter we present three models that compute the charged hadron nuclear modification factor for \oooo{} and \nene{} collisions at the LHC.
    The first is the Simple model based on BDMPS-Z in the multiple soft limit with the harmonic oscillator approximation. Second, we use a phenomenological model extrapolating experimental results for PbPb (central) collisions to other heavy and light ion collisions. Finally, we use the interface of the \emph{Trajectum} hydrodynamic medium with the Monte Carlo jet evolution code \jewel. 
    % \ik{CMS data here?}
    % Since these predictions were first made public, the CMS collaboration has released a measurement of the charged particle nuclear modification factor \cite{CMS:2025bta}.
    
    We show all three of these predictions using nuclear structure input   
    as computed from the framework of Nuclear Lattice Effective Field Theory (NLEFT, \cite{Lee:2008fa,Lahde:2019npb,Lee:2020meg}) simulations as well as the the ab initio Projected Generator Coordinate Method (PGCM, \cite{Frosini:2021tuj,Yao:2018qjv,Yao:2019rck,Frosini:2021fjf,Frosini:2021sxj,Frosini:2021ddm}) for \oo{} and \nee{}.
    For \pb{} we take neutron skin parameters from \cite{Nijs:2025qxm} (with neutron skin thickness $a_n = 0.66$) and for \xe{} the standard elliptic deformation of $\beta_2 = 0.202$ (without any triaxiality for this study).
  %  For \pb{} and \xe{} predictions we rely on \trento{} initial conditions\ik{cite}.
    The path-length and JEWEL implementations additionally make use of the hydrodynamic framework \emph{Trajectum} \cite{Nijs:2020ors,Nijs:2020roc}.
    \textit{Trajectum} is a state-of-the-art heavy-ion simulation framework that combines models of the initial conditions, pre-equilibrium evolution, viscous hydrodynamics, and hadronic transport. 
    Its parameters are constrained through Bayesian analyses of a wide set of experimental observables in lead-lead collisions, ensuring a quantitatively reliable description of the medium produced in heavy-ion collisions. 
    A table summarising the systematic effects taken into account can be found in the Supplemental Material (SM, including reference \cite{Andres:2022bql}).
    
    We will first detail the three models, and then end with a discussion of the results. 
    A preprint of this Letter appeared before light ion measurements were done and hence are true predictions. At the final submission the data is now available \cite{CMS:2025bta} and we will make a brief comment in the discussion.

\section{The Simple model \cite{Huss:2020whe}\label{sec:simple}}

    Despite its name, the Simple model is not all that simple. In practice it uses the so-called Simple formula due to Peter Arnold \cite{Arnold:2008iy}, which solves the full BDMPS-Z \cite{Baier:1996kr,Zakharov:1996fv} emission equations in the harmonic oscillator approximation for a dynamic medium. It moreover includes realistic quark and gluon spectra, realistic fragmentation functions and, in this work, EPPS16 nuclear PDFs \cite{Eskola:2016oht}. The geometry is relatively simple and boost invariant, but matches Trento \cite{Moreland:2014oya} calculations in anisotropy and in size both for the plasma as well as for the locations of the binary collisions. 
    
\begin{figure}[t]
            \centering
            \includegraphics[width=8.5cm]{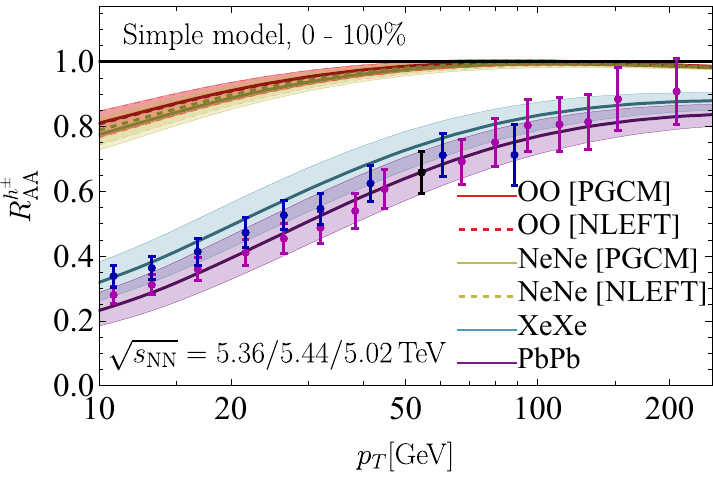}\\
            \includegraphics[width=8.5cm]{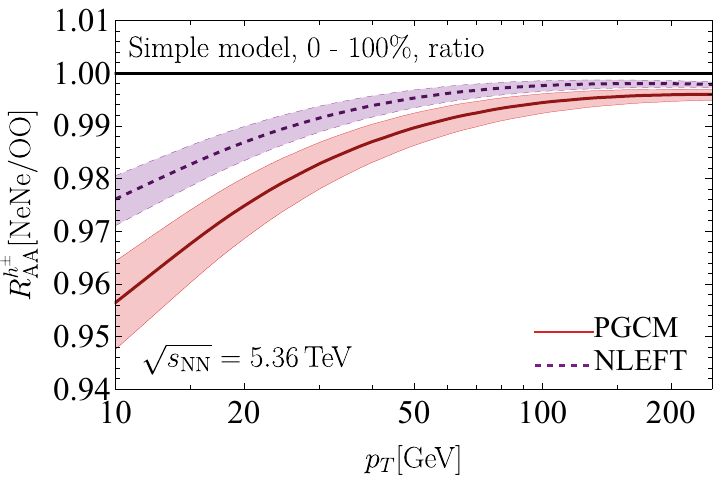}\\
            \caption{(top) Minimum bias charged
            hadron nuclear modification factor as a function of $p_{T}$ for all
            ion species collided at the LHC as predicted by the Simple model \cite{Huss:2020whe}, compared to data from CMS \cite{CMS:2016xef}. 
            (bottom) Ratio of charged hadron nuclear modification factor for \nee{} over that for \oo{}, for both PGCM and      NLEFT nuclear structure profiles.\label{fig:simpleraa}}
        \end{figure}

    Crucially, the model contains only one free parameter ($\hat{q}$) that is fitted to minimum bias charged hadron $R_{AA}$ at 54 GeV in PbPb collisions at $\sqrt{s_{NN}}=5.02$ TeV (black datapoint in top panel of \cref{fig:simpleraa}). Both the transverse momentum ($p_T$) and the centrality dependence are then a prediction of the model.
    
    One major omission in \cite{Huss:2020whe} is the absence of \nee{} as a light ion system. Therefore we took this opportunity to redo the calculation, but now for updated Trento profiles for both \oo{} and \nee{}, both from PGCM and NLEFT. The results of this exercise can be seen in \cref{fig:simpleraa}.
    
    Perhaps unsurprisingly the difference between both \oo{} and \nee{}, as well as the two structure calculations, is within the uncertainty (which, as in \cite{Huss:2020whe}, is taken to be only the uncertainty propagated to $\hat{q}$ from the CMS datapoint). Nevertheless, in the ratio this uncertainty cancels almost completely, and we see both a higher energy loss in \nee{} as well as a clear difference between PGCM and NLEFT. We will come back to this difference in the Discussion.
    
\section{A temperature-weighted path approach \cite{vanderSchee:2023uii}\label{sec:path-length}}

In this section, we estimate the nuclear modification factor ($R_{AA}$) by assuming
a power-law spectrum ($\sigma_{pp}\propto\ensuremath{p_{T}^{-6}}$) and a $p_T$-dependent energy loss ($\delta e(p_{T})$) that depends on
the temperature-weighted integral $\mathcal{I}$ according to \cite{vanderSchee:2023uii}:
    % \begin{align}
    % R_{AA} & = \sigma_{pp}(p_{T}+\delta e(p_{T}))/\sigma_{pp}(p_{T}),\label{eq:path-length}\\
    % \delta e(p_{T}) & =\kappa(p_{T})\int T^{3}{\bf {u}\cdot{\bf {dL}}},\nonumber 
    % \end{align}
    \begin{align}
    R_{AA} & = \sigma_{pp}(p_{T}+\delta e(p_{T}))/\sigma_{pp}(p_{T}),\label{eq:path-length}\\
    \delta e(p_{T}) & =\kappa(p_{T})\, \mathcal{I},\quad \mathcal{I} \equiv \int T^{3}{\bf {u}\cdot{\bf {dL}}},\nonumber 
    \end{align}
where we assume that the proportionality constant $\kappa(p_{T})$ depends only on $p_{T.}$ 
Within this setup, we fix $\kappa(p_{T})$ from an experimental $R^{\text{ref}}_{AA}(p_T)$ \cite{CMS:2016xef} and path $\mathcal{I}^\text{ref}$ in a reference centrality class (propagating its total uncertainty). We can determine any $R_{AA}(p_{T})$ when given the temperature-weighted path $\mathcal{I}$ \cite{Beattie:2022ojg}. %-length integral from \emph{Trajectum} \cite{Beattie:2022ojg}. 
In this work we choose the 0-5\% centrality class, % to fit $\kappa(p_{T})$, 
since this gives the most accurate value. Nevertheless, in the region of interest ($p_T > 5\,$GeV) any centrality class would have given almost identical results (albeit with a larger uncertainty).
%Solving $\kappa(p_{T})$ can be done  using a single centrality class (e.g. PbPb 0-5\%), but since we are interested in estimating the light ion $R_{AA}$ we will here use all five centrality classes up to 70\% that are measured by CMS \cite{CMS:2016xef} and treat the spread as a systematic uncertainty. 
Secondly, we take 10 different
parameter settings from the posterior in \emph{Trajectum} \cite{vanderSchee:2023uii} and treat
the spread as a second systematic uncertainty\footnote{Note that it is important to use the same settings when taking ratios or making predictions; those are determined for each setting individually and the uncertainty is obtained as a final step.}.

It is interesting that, since we fix $\kappa(p_T)$ from the experimental $R^{\text{ref}}_{AA}$, our results are rather insensitive to the assumed $\sigma_{pp}(p_T)$. Indeed, expanding in the power $p$ gives
    \begin{equation}
    R_{AA}  = (R^{\text{ref}}_{AA})^{\mathcal{I} / \mathcal{I}^\text{ref}}
    \left(1+\frac{\mathcal{I}(\mathcal{I}-\mathcal{I}^\text{ref})\log(R^{\text{ref}}_{AA})^2}{2(\mathcal{I}^\text{ref})^2 p}+\mathcal{O}(p^2)\right).\label{eq:path-length-expansion}
    \end{equation}
We now see that for any $\mathcal{I}/\mathcal{I}^\text{ref}\in [0,1]$ and $R^{\text{ref}}_{AA} > 0.25$ the subleading term is smaller than 4\%, which explains the insensitivity to the $pp$ spectrum.

\begin{figure}
        \includegraphics[width=8cm]{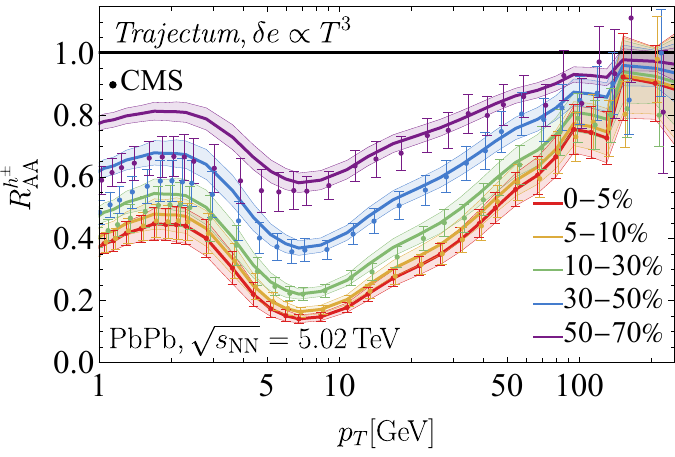}
        \caption{PbPb charged hadron nuclear
        modification factor as a function of $p_{T}$ and centrality.
        The combination of the other centralities provides the basis to postdict
        a given centrality according to \cref{eq:path-length}.\label{fig:RAAPb}}
    \end{figure}

For the purposes of validation, in \cref{fig:RAAPb} we present the nuclear modification
factor for the five different centrality classes.%, here excluding that particular centrality class used for the fit. 
We deliberately
also show the $p_{T}$ range below 5 GeV, where \cref{eq:path-length}
is not expected to apply since this starts to enter the hydrodynamic regime.
Above this range the centrality and
$p_{T}$ dependence is captured remarkably well. 
The spread in the curves is the combination of experimental uncertainty and the spread due to the \emph{Trajectum} posterior uncertainty.
%the \emph{Trajectum} posterior uncertainty and the spread due to varying the centrality class that is used for the initial fit, but does not include the experimental uncertainty.
% Since there is such a good agreement above 5 GeV it is in fact hard to distinguish the result from the one if one would have used a single centrality class.
%Above 5 GeV, the narrowness of the uncertainty band is indicative of the fact that the prediction is not sensitive to the choice of centrality class used for the initial fit.
%\ik{Is the \emph{Trajectum} posterior uncertainty just a flat number across all curves? It doesn't depend on the centrality?}\ws{hmm.. it's a bit more subtle I guess, since it's the sum of varying parameters and varying centrality class. But I think the sentence is good enough :)}

\begin{figure}
        \includegraphics[width=8cm]{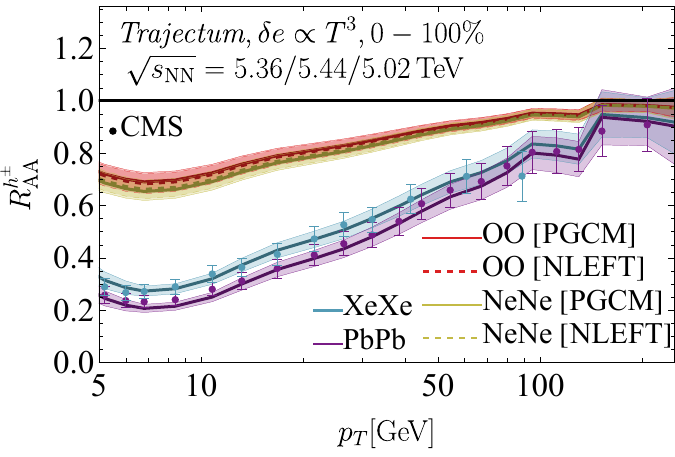}
        \includegraphics[width=8cm]{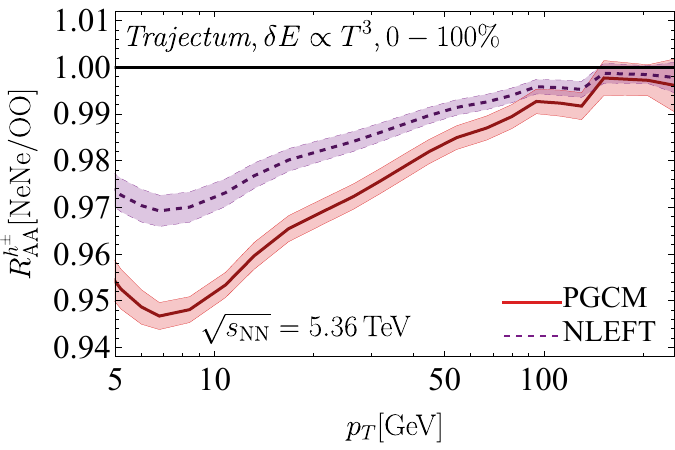}
        \caption{(top) Minimum bias path-length approach to the charged
        hadron nuclear modification factor as a function of $p_{T}$ for all
        ion species collided at the LHC. 
        (bottom) Ratio of charged hadron nuclear modification factor for \nee{} over that for \oo{}, for both PGCM and      NLEFT nuclear structure profiles.\label{fig:RAAMB}}
    \end{figure}

Now that we have validated our rather basic model, we show in
\cref{fig:RAAMB} predictions for \oooo{} and \nene{} collisions (top),
together with the \nene{} over \oooo{} ratio (bottom), again for both
PGCM and NLEFT nuclear structure predictions. In the main plot the
uncertainties make it difficult to distinguish \oo{} from \nee{}
or PGCM from NLEFT, but looking more carefully at the ratio, it can
be seen that partons in \nee{} collisions have significantly more suppression,
and again that this effect is stronger in PGCM than in NLEFT. 
The cancellation of the uncertainties is impressive, but an important point that is made clear in the bottom panel of \cref{fig:RAAMB}, is that the difference between PGCM and NLEFT is a systematic uncertainty in and of itself.
The fact that NLEFT has a smaller difference between \oo{} and \nee{} is a somewhat complicated interplay between several effects. NLEFT has a larger difference in size between the two nuclei (see also \cite{Giacalone:2024luz}), which in turn implies a smaller difference in the initial QGP temperature. It turns out that the temperature-weighted paths differ more for PGCM than for NLEFT (see also the SM).
In anticipation of future measurements we present the 0 - 10\% centrality path-length approach in the SM.

\section{\jewel{} and \emph{Trajectum} \cite{Kolbe:2023rsq} \label{sec:jeweltrajectum}}

    \jewel{} (Jet Evolution With Energy Loss) \cite{Zapp:2012ak,Zapp:2013vla,Zapp:2008gi} is a Monte Carlo event generator designed to model parton showers in the presence of a QCD medium. 
    It incorporates both elastic and inelastic interactions between the jet and medium constituents, and includes a treatment of medium recoil, allowing for studies of the backreaction of the plasma and jet-medium correlations. 
    In its standard configuration, \jewel{} is coupled to a simplified, boost-invariant medium model in which overlapping nuclear density distributions are used in conjunction with the equation of state for an ideal gas of quarks and gluons and longitudinal expansion.

    A publicly available interface for \jewel{} \cite{Kolbe:2023rsq} connects \jewel{} to medium profiles generated by \textit{Trajectum}.
    The output of \textit{Trajectum} can consist of event-by-event hydrodynamic profiles of energy density, temperature, and flow, which can be used as realistic backgrounds for hard probes.
    Instead of relying on the built-in medium model, hard partons can now evolve through dynamically expanding, event-by-event fluctuating backgrounds with spatially resolved temperature and flow fields. 
    This allows a consistent treatment of parton-medium interactions across different collision systems, including light-ion collisions where fluctuations in the medium are expected to be dominant. 
    The interface does not change any of the core \jewel{} code, separating the medium modelling entirely from the partonic evolution and sampling only temperature and velocity values.
    By embedding hard partons from \jewel{} in these realistic environments, the interface extends \jewel's applicability to a broader range of systems and observables, enabling systematic studies of energy loss in evolving media.

\begin{figure}
    \includegraphics[width=8cm]{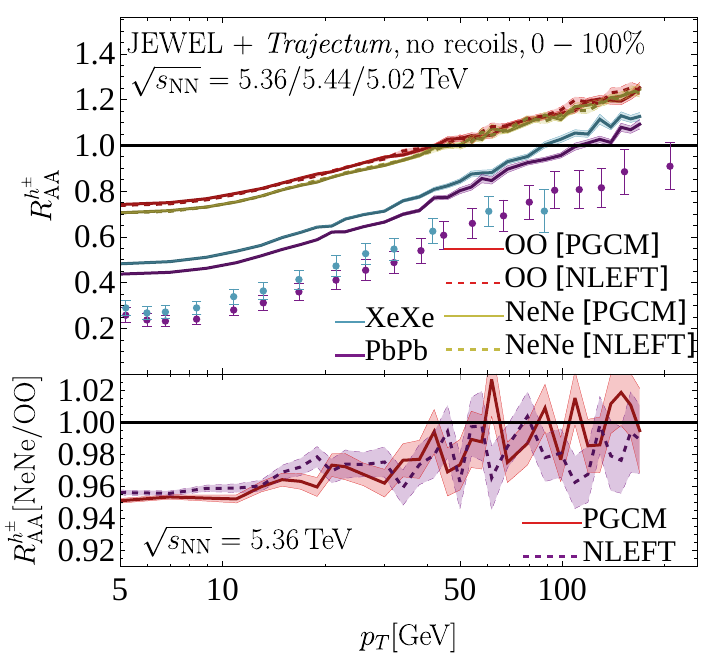}
    \caption{The charged hadron nuclear modification factor as a function of $p_{T}$ for all
    ion species collided at the LHC as computed with the JEWEL-Trajectum interface  (uncertainties are statistical only). \label{fig:jewelraaMB}}
    \end{figure}

    We present the charged hadron nuclear modification factor in  \cref{fig:jewelraaMB}. 
    The results presented here are untuned in the sense that default parameters were used wherever possible: 
    For \emph{Trajectum} these are from \cite{Nijs:2025qxm}. 
    \jewel{} version 2.4.0 was used with default values for all the parameters in the core \jewel{} code, except for using EPPS21 nuclear PDFs \cite{Eskola:2021nhw}. 
    We used the same \oo{} nPDF set for both \oo{} and \nee{} since no dedicated \nee{} set is available.
    The nPDFs (and their uncertainties) pose a limit on accuracy that is universal across energy-loss computations in \oo{} and \nee{} as they are not well-known.
    % The full list of \jewel parameters used in the present letter is shown in \cref{table:\jewelParams}; 

    %As the name suggests, \jewel{} is better suited to describe jets than charged hadrons. As such, it does not describe the charged hadron spectrum for Pb or Xe collisions, although it should be noted that, unlike \cref{sec:simple,sec:path-length}, we did not tune \jewel{} here. 
    \jewel{} underpredicts the energy loss for hadrons, with a noteworthy rise above unity. 
    For the most part, this feature is the result of a kinematical effect in which the elastic scatterings convert longitudinal momentum into transverse momentum.
    At the same time, the energy-loss effect disappears at high-$p_T$. As a result, the $p_T$ spectrum is hardened, as was already found in \cite{PHENIX:2015siv} (see Fig. 1.33) and \cite{Zapp:2012ak}. 
    In jets, the effect is somewhat mitigated because much of this transverse momentum is carried outside of the jet cone. 
    In the SM we present the same \jewel{} results, but for the $R=0.4$ anti-$k_T$ jet nuclear modification,  compared to ATLAS \cite{ATLAS:2018gwx} and central ALICE \cite{ALICE:2023waz} data. 
    For ATLAS, \jewel{} overpredicts the energy loss while there is good agreement with central ALICE data. This makes it unlikely that we can fully tune \jewel{} even for Xe and Pb collisions.
%    \ik{Compare also to 0-10\% ALICE jets, then argue it isn't too bad, especially considering that there is no tuning at all.}
    %Note that there is $R=0.2$ charged jet  $R_\text{AA}$ data available in \cite{ALICE:2023waz}, but not for $0 - 100\%$ centrality. Nevertheless, the JEWEL nuclear modification factor for jets is not far off.
    Nevertheless, quite pleasingly, we see that, although the $p_T$-dependence is different from the other models, the ion-size dependence is qualitatively similar, with the \nene/\oooo{} ratio even quantitatively so.

%Perhaps naively, one may have expected \jewel{} to have similar differences between PGCM and NLEFT since we use the same hydrodynamics profiles as in Section \ref{sec:path-length}. 
Curiously, as opposed to Fig.~\ref{fig:RAAMB}, \jewel{} finds almost no difference between PGCM and NLEFT in the ratio in Fig.~\ref{fig:jewelraaMB}, even when using the same hydrodynamic profiles as in Section \ref{sec:path-length}. 
It is important to stress that \jewel{} is considerably richer than the path based approach and, in particular, contains fluctuations in the jet fragmentation that is known to be crucial in e.g. the dijet asymmetry \cite{Milhano:2015mng}. These fluctuations dominate or interplay in a non-trivial way with the temperature and plasma size such that, in the end, the PGCM and NLEFT ratios turn out to be almost identical.

\section*{Discussion}

   \begin{figure}[t]
        \centering
        \includegraphics[width=8.5cm]{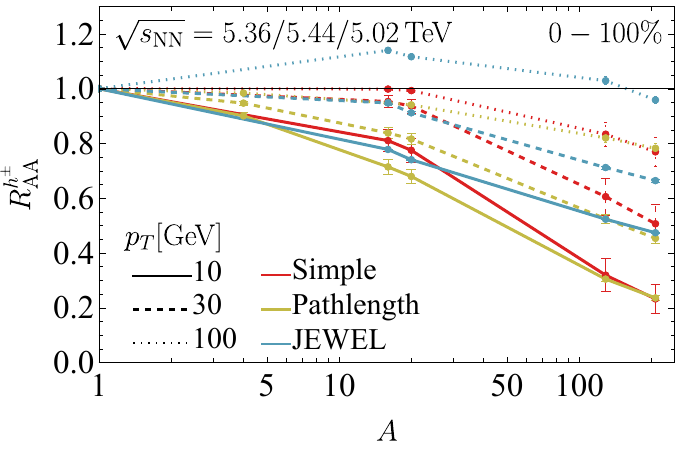}\\
        \caption{Minimum bias charged
        hadron nuclear modification factor as a function of ion number for several $p_T$ values and the three models studied in this Letter.
        \label{fig:finalplot}}
    \end{figure}

    The most interesting result of this letter is that the bands of the \oo{} and \nee{} nuclear modification factors for the
    path-length approach of \cref{sec:path-length} are significantly lower than those of the Simple
    model of \cref{sec:simple} despite the fact that both are fitted to the same lead data. 
    At least in hindsight this should not be surprising. 
    In the multiple soft scattering approximation, energy loss starts much slower
    due to coherence effects (see for instance the H.O. approximation in Fig 2 of \cite{Caron-Huot:2010qjx} for
    a clarifying explanation). 
    Since both approaches fit the minimum bias
    nuclear modification factor, it should be expected that, for small systems,
    energy loss is estimated to be smaller in a BDMPS-Z-based approach than in
    a path-length approach which does not include such coherence (in Fig
    2 of \cite{Caron-Huot:2010qjx} it would be closer to AMY). 
    The fact that the 50-70\%
    centrality bin (similar in size as 0-10\% light ion collisions) is
    better-estimated by the path-length approach than by the Simple model
    (compare \cref{fig:RAAPb} with fig. 8 in \cite{Huss:2020whe}) suggests that the experimental measurement
    could be closer to the path-length approach.
    The JEWEL results are more difficult to interpret as they do not fit the lead data for charged hadrons.
    The fit to jet data is better (see SM).
    Nevertheless, it is clear that even the JEWEL charged hadron result exhibits the expected system-size dependence.

    A second interesting outcome is that the \nene/\oooo{}
    ratio is significantly different for PGCM and NLEFT, by about a factor
    two when looking at $(R_{AA}-1)$ in the path based approach.
    In hindsight, this too could have
    been anticipated by the mean transverse momentum, as presented in \cite{Giacalone:2024luz}, where the PGCM and NLEFT also differ by an amount much larger than the
    systematic uncertainty. 
    The mean transverse
    momentum is directly related to the size of the initial QGP and indeed
    the \nene/\oooo{} ratio of the size could explain this
    difference (see also the SM). When comparing to the (accurately known) charge radii
    of \oo{} and \nee{} it turned out that both PGCM and NLEFT were off
    by about the same (but opposite sign) amount \cite{Giacalone:2024luz}, such that it is likely
    that the true ratio of %both the mean transverse momentum as well as
    the $R_{AA}$ ratio will lie somewhere in the middle \footnote{Experimentally it may also be challenging to obtain such a precise ratio; many systematic uncertainties likely cancel, but the luminosity uncertainty is sizeable and substantially uncorrelated between the two light ion runs.}.
  %  \ik{Is the mean transverse momentum and charge radius really the important thing to discuss here? Aren't the temperature and pathlength the more important ones? }
   % \ik{Is the data likely to be precise enough here to discriminate?}
    
    Arguably, our study is not complete. Section \ref{sec:path-length} does not include the
    uncertainty coming from the experimental data, but includes the systematic
    uncertainty from \emph{Trajectum} and from the choice of centrality class to gauge
    the model to. \Cref{sec:simple} on the other hand only includes the uncertainty of
    the single data point that the model is gauged to. The nPDF uncertainties
    further complicate the predictions significantly and are, for example, not
    included at all in \cref{sec:path-length}. In the ratio $\text{NeNe}/\text{OO}$
    those uncertainties likely largely cancel, but it is not even known
    if nPDF modification is monotonic with nucleus size (in fact, one
    could argue that \nee{} is more like an \oo{} plus $^{4}\text{He}$ and might
    inherit nPDFs a bit from both). 
    As such, we did not attempt to include
    such effects or, even more challenging, their (correlated) uncertainties (see however \cite{Mazeliauskas:2025clt, Jonas:2026yoz}).
    
    Our results are nicely summarised in \cref{fig:finalplot}. 
    Here we show, for several values of $p_T$, how the hadron $R_\text{AA}$ depends on the ion species for the three models considered. 
    The non-trivial path length dependence of all these models can clearly be seen, especially for the Simple model.
    For the pathlength model we also computed the $R_{\rm AA}^{h\pm}$ for $A=4$ ($^4$He$^4$He). Even for $p_T=10\,$GeV the energy loss is small and will be challenging (though exciting) to probe given the nPDF uncertainties.
    
    %For the reasons described above, we also show the result for jets in \jewel. The non-trivial path length dependence of all these models can clearly be seen, especially for the Simple model. However, a precise answer to questions like ``how small can the atomic mass number be before energy-loss effects become negligible?'' remains elusive: \ik{On the one hand, the main theoretical limitation, for the time-being, lies in the nPDF uncertainties, which may suggest an answer along the lines of ``negligable compared to what?'' \cite{Faraday:2025prr}, but as a community we remain hopeful that this is, in principle, an easily surmountable limitation. On the other hand, many of the smallest systems are asymmetric, causing the question to be less well-posed. In a future work we will be exploring the pathlength and system-size dependence of a wider variety of collision systems in greater detail. Nevertheless, we simply quote here the }
    % since none of the three models take all systematic uncertainties fully into account, with nPDF uncertainties posing an acute difficulty.
    % \ik{Should we say something here about nPDF uncertainties for HeHe for instance being on the order of 10\%, of similar magnitude as the energy-loss?}
    
    All-in-all there are exciting times ahead. Not only to see parton
    energy loss in small systems, but also to distinguish between simple
    and advanced models that can teach us more about the path length dependence
    of energy loss. It is interesting how much these high $p_{T}$ results
    are linked to the soft sector and understanding the shape of the ions
    or the resulting QGP.
    
    At the final submission, the experimental light-ion data is now available \cite{CMS:2025bta}, which 
    gives an excellent match to the pathlength approach (see Fig.~3 in \cite{CMS:2025bta}). Both the simple model and JEWEL underpredict the energy loss. In hindsight, this could perhaps be expected for the simple model, since this model also underpredicts the energy loss in the 50-70\% centrality class for PbPb (see Fig.~8 in \cite{Huss:2020whe}). This could indicate that BDMPS-Z energy loss (which scales as $L^2$ for small lengths) has too little energy loss for small droplets of QGP.

    The data that support the findings of this article are publicly available \cite{van_der_schee_2025_22257566}.

    {\bf Acknowledgements. }
    We thank Gian Michele Innocenti, Aleksas Mazeliauskas for interesting discussions and Gian Michele especially for suggesting \cref{fig:finalplot}.
    This work is supported by the DST/NRF in South Africa under Thuthuka grant number TTK240313208902.
    
    %\section{Results and Discussion\label{sec:results}}

\bibliographystyle{apsrev4-1}
\bibliography{main1}

@article{Citron:2018lsq,
    author = "Citron, Z. and others",
    editor = "Dainese, Andrea and Mangano, Michelangelo and Meyer, Andreas B. and Nisati, Aleandro and Salam, Gavin and Vesterinen, Mika Anton",
    title = "{Report from Working Group 5}: {Future physics opportunities for high-density QCD at the LHC with heavy-ion and proton beams}",
    eprint = "1812.06772",
    archivePrefix = "arXiv",
    primaryClass = "hep-ph",
    reportNumber = "CERN-LPCC-2018-07",
    doi = "10.23731/CYRM-2019-007.1159",
    journal = "CERN Yellow Rep. Monogr.",
    volume = "7",
    pages = "1159--1410",
    year = "2019"
}

@article{Nijs:2025qxm,
    author = "Nijs, Govert and van der Schee, Wilke",
    title = "{Transmutation of 16O and 20Ne at the Large Hadron Collider}",
    eprint = "2507.01659",
    archivePrefix = "arXiv",
    primaryClass = "nucl-th",
    reportNumber = "CERN-TH-2025-127",
    doi = "10.1016/j.physletb.2025.140061",
    journal = "Phys. Lett. B",
    volume = "872",
    pages = "140061",
    year = "2026"
}

@article{vanderSchee:2023uii,
    author = "van der Schee, Wilke and Lee, Yen-Jie and Nijs, Govert and Chen, Yi",
    title = "{Hard probes in isobar collisions as a probe of the neutron skin}",
    eprint = "2307.11836",
    archivePrefix = "arXiv",
    primaryClass = "nucl-th",
    reportNumber = "CERN-TH-2023-140, MIT-CTP/5588",
    doi = "10.1016/j.physletb.2024.138953",
    journal = "Phys. Lett. B",
    volume = "856",
    pages = "138953",
    year = "2024"
}

@article{Giacalone:2024luz,
    author = "Giacalone, Giuliano and others",
    title = "{Exploiting Ne20 Isotopes for Precision Characterizations of Collectivity in Small Systems}",
    eprint = "2402.05995",
    archivePrefix = "arXiv",
    primaryClass = "nucl-th",
    reportNumber = "CERN-TH-2024-021",
    doi = "10.1103/k8rb-jgvq",
    journal = "Phys. Rev. Lett.",
    volume = "135",
    number = "1",
    pages = "012302",
    year = "2025"
}

@article{Giacalone:2024ixe,
    author = "Giacalone, Giuliano and others",
    title = "{Anisotropic Flow in Fixed-Target Pb208+Ne20 Collisions as a Probe of Quark-Gluon Plasma}",
    eprint = "2405.20210",
    archivePrefix = "arXiv",
    primaryClass = "nucl-th",
    reportNumber = "CERN-TH-2024-074",
    doi = "10.1103/PhysRevLett.134.082301",
    journal = "Phys. Rev. Lett.",
    volume = "134",
    number = "8",
    pages = "082301",
    year = "2025"
}

@article{Summerfield:2021oex,
    author = "Summerfield, Nicholas and Lu, Bing-Nan and Plumberg, Christopher and Lee, Dean and Noronha-Hostler, Jacquelyn and Timmins, Anthony",
    title = "{$^{16}$O $^{16}$O collisions at energies available at the BNL Relativistic Heavy Ion Collider and at the CERN Large Hadron Collider comparing $\alpha$ clustering versus substructure}",
    eprint = "2103.03345",
    archivePrefix = "arXiv",
    primaryClass = "nucl-th",
    doi = "10.1103/PhysRevC.104.L041901",
    journal = "Phys. Rev. C",
    volume = "104",
    number = "4",
    pages = "L041901",
    year = "2021"
}

@article{YuanyuanWang:2024sgp,
    author = "Wang, Yuanyuan and Zhao, Shujun and Cao, Boxing and Xu, Hao-jie and Song, Huichao",
    title = "{Exploring the compactness of {\ensuremath{\alpha}} clusters in O16 nuclei with relativistic O16+O16 collisions}",
    eprint = "2401.15723",
    archivePrefix = "arXiv",
    primaryClass = "nucl-th",
    doi = "10.1103/PhysRevC.109.L051904",
    journal = "Phys. Rev. C",
    volume = "109",
    number = "5",
    pages = "L051904",
    year = "2024"
}

@inproceedings{Brewer:2021kiv,
    author = "Brewer, Jasmine and Mazeliauskas, Aleksas and van der Schee, Wilke",
    title = "{Opportunities of OO and $p$O collisions at the LHC}",
    booktitle = "{Opportunities of OO and pO collisions at the LHC}",
    eprint = "2103.01939",
    archivePrefix = "arXiv",
    primaryClass = "hep-ph",
    reportNumber = "CERN-TH-2021-028",
    month = "3",
    year = "2021"
}

@inproceedings{Huang:2023viw,
    author = "Huang, Shengli",
    title = "{Measurements of azimuthal anisotropies in $^{16}$O+$^{16}$O and $γ$+Au collisions from STAR}",
    eprint = "2312.12167",
    archivePrefix = "arXiv",
    primaryClass = "nucl-ex",
    month = "12",
    year = "2023"
}

@article{ATLAS:2025nnt,
    author = "Aad, Georges and others",
    collaboration = "ATLAS",
    title = "{Measurement of the azimuthal anisotropy of charged particles in $\sqrt{s_{\mathrm{NN}}}=5.36$ TeV $^{16}$O$+^{16}$O and $^{20}$Ne$+^{20}$Ne collisions with the ATLAS detector}",
    eprint = "2509.05171",
    archivePrefix = "arXiv",
    primaryClass = "nucl-ex",
    reportNumber = "CERN-EP-2025-200",
    month = "9",
    year = "2025"
}

@article{ALICE:2025luc,
    author = "Abualrob, Ibrahim Jaser and others",
    collaboration = "ALICE",
    title = "{Evidence of nuclear geometry-driven anisotropic flow in OO and Ne$-$Ne collisions at $\mathbf{\sqrt{{\textit s}_{\rm\mathbf {NN}}}}$ = 5.36 TeV}",
    eprint = "2509.06428",
    archivePrefix = "arXiv",
    primaryClass = "nucl-ex",
    reportNumber = "CERN-EP-2025-203",
    month = "9",
    year = "2025"
}

@article{CMS:2025tga,
    author = "Hayrapetyan, Aram and others",
    collaboration = "CMS",
    title = "{Observation of long-range collective flow in OO and NeNe collisions and implications for nuclear structure studies}",
    eprint = "2510.02580",
    archivePrefix = "arXiv",
    primaryClass = "nucl-ex",
    reportNumber = "CMS-HIN-25-009, CERN-EP-2025-222",
    month = "10",
    year = "2025"
}

@article{STAR:2025ivi,
    collaboration = "STAR",
    title = "{Engineering the shapes of quark-gluon plasma droplets by comparing anisotropic flow in small symmetric and asymmetric collision systems}",
    eprint = "2510.19645",
    archivePrefix = "arXiv",
    primaryClass = "nucl-ex",
    month = "10",
    year = "2025"
}

@article{Grosse-Oetringhaus:2024bwr,
    author = "Grosse-Oetringhaus, Jan Fiete and Wiedemann, Urs Achim",
    title = "{A Decade of Collectivity in Small Systems}",
    eprint = "2407.07484",
    archivePrefix = "arXiv",
    primaryClass = "hep-ex",
    reportNumber = "CERN-TH-2024-110",
    month = "7",
    year = "2024"
}

@article{CMS:2010ifv,
    author = "Khachatryan, Vardan and others",
    collaboration = "CMS",
    title = "{Observation of Long-Range Near-Side Angular Correlations in Proton-Proton Collisions at the LHC}",
    eprint = "1009.4122",
    archivePrefix = "arXiv",
    primaryClass = "hep-ex",
    reportNumber = "CMS-QCD-10-002, CERN-PH-EP-2010-031",
    doi = "10.1007/JHEP09(2010)091",
    journal = "JHEP",
    volume = "09",
    pages = "091",
    year = "2010"
}

@article{ALICE:2016fzo,
    author = "Adam, Jaroslav and others",
    collaboration = "ALICE",
    title = "{Enhanced production of multi-strange hadrons in high-multiplicity proton-proton collisions}",
    eprint = "1606.07424",
    archivePrefix = "arXiv",
    primaryClass = "nucl-ex",
    reportNumber = "CERN-EP-2016-153",
    doi = "10.1038/nphys4111",
    journal = "Nature Phys.",
    volume = "13",
    pages = "535--539",
    year = "2017"
}

@article{ALICE:2013snh,
    author = "Abelev, Betty Bezverkhny and others",
    collaboration = "ALICE",
    title = "{$J/\psi$ production and nuclear effects in p-Pb collisions at $\sqrt{S_{NN}}$ = 5.02 TeV}",
    eprint = "1308.6726",
    archivePrefix = "arXiv",
    primaryClass = "nucl-ex",
    reportNumber = "CERN-PH-EP-2013-163, ALICE-PUBLIC-2013-002",
    doi = "10.1007/JHEP02(2014)073",
    journal = "JHEP",
    volume = "02",
    pages = "073",
    year = "2014"
}

@article{ALICE:2014cgk,
    author = "Abelev, Betty Bezverkhny and others",
    collaboration = "ALICE",
    title = "{Suppression of $\psi$(2S) production in p-Pb collisions at $\sqrt{s_{\rm NN}}$ = 5.02 TeV}",
    eprint = "1405.3796",
    archivePrefix = "arXiv",
    primaryClass = "nucl-ex",
    reportNumber = "CERN-PH-EP-2014-092",
    doi = "10.1007/JHEP12(2014)073",
    journal = "JHEP",
    volume = "12",
    pages = "073",
    year = "2014"
}

@article{ALICE:2020msa,
    author = "Acharya, Shreyasi and others",
    collaboration = "ALICE",
    title = "{Multiplicity dependence of J/$\psi$ production at midrapidity in pp collisions at $\sqrt{s}$ = 13 TeV}",
    eprint = "2005.11123",
    archivePrefix = "arXiv",
    primaryClass = "nucl-ex",
    reportNumber = "CERN-EP-2020-088",
    doi = "10.1016/j.physletb.2020.135758",
    journal = "Phys. Lett. B",
    volume = "810",
    pages = "135758",
    year = "2020"
}

@article{Weller:2017tsr,
    author = "Weller, Ryan D. and Romatschke, Paul",
    title = "{One fluid to rule them all: viscous hydrodynamic description of event-by-event central p+p, p+Pb and Pb+Pb collisions at $\sqrt{s}=5.02$ TeV}",
    eprint = "1701.07145",
    archivePrefix = "arXiv",
    primaryClass = "nucl-th",
    doi = "10.1016/j.physletb.2017.09.077",
    journal = "Phys. Lett. B",
    volume = "774",
    pages = "351--356",
    year = "2017"
}

@article{Zhao:2017rgg,
    author = "Zhao, Wenbin and Zhou, You and Xu, Haojie and Deng, Weitian and Song, Huichao",
    title = "{Hydrodynamic collectivity in proton{\textendash}proton collisions at 13 TeV}",
    eprint = "1801.00271",
    archivePrefix = "arXiv",
    primaryClass = "nucl-th",
    doi = "10.1016/j.physletb.2018.03.022",
    journal = "Phys. Lett. B",
    volume = "780",
    pages = "495--500",
    year = "2018"
}

@article{Bierlich:2018xfw,
    author = {Bierlich, Christian and Gustafson, G{\"o}sta and L{\"o}nnblad, Leif and Shah, Harsh},
    title = "{The Angantyr model for Heavy-Ion Collisions in PYTHIA8}",
    eprint = "1806.10820",
    archivePrefix = "arXiv",
    primaryClass = "hep-ph",
    reportNumber = "LU-TP-18-19, LU-TP 18-19, MCnet-18-12",
    doi = "10.1007/JHEP10(2018)134",
    journal = "JHEP",
    volume = "10",
    pages = "134",
    year = "2018"
}

@article{Huss:2020dwe,
    author = "Huss, Alexander and Kurkela, Aleksi and Mazeliauskas, Aleksas and Paatelainen, Risto and van der Schee, Wilke and Wiedemann, Urs Achim",
    title = "{Discovering Partonic Rescattering in Light Nucleus Collisions}",
    eprint = "2007.13754",
    archivePrefix = "arXiv",
    primaryClass = "hep-ph",
    reportNumber = "CERN-TH-2020-126",
    doi = "10.1103/PhysRevLett.126.192301",
    journal = "Phys. Rev. Lett.",
    volume = "126",
    number = "19",
    pages = "192301",
    year = "2021"
}

@article{Faraday:2024qtl,
    author = "Faraday, Coleridge and Horowitz, W. A.",
    title = "{A unified description of small, peripheral, and large system suppression data from pQCD}",
    eprint = "2411.09647",
    archivePrefix = "arXiv",
    primaryClass = "hep-ph",
    doi = "10.1016/j.physletb.2025.139437",
    journal = "Phys. Lett. B",
    volume = "864",
    pages = "139437",
    year = "2025"
}

@article{ALICE:2016faw,
    author = "Adam, Jaroslav and others",
    collaboration = "ALICE",
    title = "{Centrality dependence of charged jet production in p-Pb collisions at $\sqrt{s_\mathrm{NN}}$ = 5.02 TeV}",
    eprint = "1603.03402",
    archivePrefix = "arXiv",
    primaryClass = "nucl-ex",
    reportNumber = "CERN-EP-2016-052, ALICE-PUBLIC-2016-001",
    doi = "10.1140/epjc/s10052-016-4107-8",
    month = "3",
    year = "2016"
}

@article{ALICE:2017svf,
    author = "Acharya, Shreyasi and others",
    collaboration = "ALICE",
    title = "{Constraints on jet quenching in p-Pb collisions at $\mathbf{\sqrt{s_{NN}}}$ = 5.02 TeV measured by the event-activity dependence of semi-inclusive hadron-jet distributions}",
    eprint = "1712.05603",
    archivePrefix = "arXiv",
    primaryClass = "nucl-ex",
    reportNumber = "CERN-EP-2017-324",
    doi = "10.1016/j.physletb.2018.05.059",
    journal = "Phys. Lett. B",
    volume = "783",
    pages = "95--113",
    year = "2018"
}

@article{ALICE:2014xsp,
    author = "Adam, Jaroslav and others",
    collaboration = "ALICE",
    title = "{Centrality dependence of particle production in p-Pb collisions at $\sqrt{s_{\rm NN} }$= 5.02 TeV}",
    eprint = "1412.6828",
    archivePrefix = "arXiv",
    primaryClass = "nucl-ex",
    reportNumber = "CERN-PH-EP-2014-281",
    doi = "10.1103/PhysRevC.91.064905",
    journal = "Phys. Rev. C",
    volume = "91",
    number = "6",
    pages = "064905",
    year = "2015"
}

@article{ATLAS:2022kqu,
    author = "Aad, Georges and others",
    collaboration = "ATLAS",
    title = "{Charged-hadron production in $pp$, $p$+Pb, Pb+Pb, and Xe+Xe collisions at $\sqrt{s_{_\text{NN}}}=5$ TeV with the ATLAS detector at the LHC}",
    eprint = "2211.15257",
    archivePrefix = "arXiv",
    primaryClass = "hep-ex",
    reportNumber = "CERN-EP-2022-221",
    doi = "10.1007/JHEP07(2023)074",
    journal = "JHEP",
    volume = "07",
    pages = "074",
    year = "2023"
}

@article{CMS:2016xef,
    author = "Khachatryan, Vardan and others",
    collaboration = "CMS",
    title = "{Charged-particle nuclear modification factors in PbPb and pPb collisions at $ \sqrt{s_{\mathrm{N}\;\mathrm{N}}}=5.02 $ TeV}",
    eprint = "1611.01664",
    archivePrefix = "arXiv",
    primaryClass = "nucl-ex",
    reportNumber = "CMS-HIN-15-015, CERN-EP-2016-242",
    doi = "10.1007/JHEP04(2017)039",
    journal = "JHEP",
    volume = "04",
    pages = "039",
    year = "2017"
}

@article{ALICE:2018vuu,
    author = "Acharya, S. and others",
    collaboration = "ALICE",
    title = "{Transverse momentum spectra and nuclear modification factors of charged particles in pp, p-Pb and Pb-Pb collisions at the LHC}",
    eprint = "1802.09145",
    archivePrefix = "arXiv",
    primaryClass = "nucl-ex",
    reportNumber = "CERN-EP-2018-025",
    doi = "10.1007/JHEP11(2018)013",
    journal = "JHEP",
    volume = "11",
    pages = "013",
    year = "2018"
}

@article{ALICE:2014nqx,
    author = "Abelev, Betty Bezverkhny and others",
    collaboration = "ALICE",
    title = "{Transverse momentum dependence of inclusive primary charged-particle production in p-Pb collisions at $\sqrt{s_\mathrm{{NN}}}=5.02~\text {TeV}$}",
    eprint = "1405.2737",
    archivePrefix = "arXiv",
    primaryClass = "nucl-ex",
    reportNumber = "CERN-PH-EP-2014-088",
    doi = "10.1140/epjc/s10052-014-3054-5",
    journal = "Eur. Phys. J. C",
    volume = "74",
    number = "9",
    pages = "3054",
    year = "2014"
}

@article{Lee:2008fa,
    author = "Lee, Dean",
    title = "{Lattice simulations for few- and many-body systems}",
    eprint = "0804.3501",
    archivePrefix = "arXiv",
    primaryClass = "nucl-th",
    doi = "10.1016/j.ppnp.2008.12.001",
    journal = "Prog. Part. Nucl. Phys.",
    volume = "63",
    pages = "117--154",
    year = "2009"
}

@book{Lahde:2019npb,
    author = {L{\"a}hde, Timo A. and Mei{\ss}ner, Ulf-G.},
    title = "{Nuclear Lattice Effective Field Theory}: {An introduction}",
    doi = "10.1007/978-3-030-14189-9",
    isbn = "978-3-030-14187-5, 978-3-030-14189-9",
    publisher = "Springer",
    volume = "957",
    year = "2019"
}

@article{Lee:2020meg,
    author = "Lee, Dean",
    title = "{Recent Progress in Nuclear Lattice Simulations}",
    doi = "10.3389/fphy.2020.00174",
    journal = "Front. in Phys.",
    volume = "8",
    pages = "174",
    year = "2020"
}

@article{Frosini:2021tuj,
    author = "Frosini, M. and Duguet, T. and Bally, B. and Beaujeault-Taudi{\`e}re, Y. and Ebran, J. -P. and Som{\`a}, V.",
    title = "{In-medium $k$-body reduction of $n$-body operators: A flexible symmetry-conserving approach based on the sole one-body density matrix}",
    eprint = "2102.10120",
    archivePrefix = "arXiv",
    primaryClass = "nucl-th",
    doi = "10.1140/epja/s10050-021-00458-z",
    journal = "Eur. Phys. J. A",
    volume = "57",
    number = "4",
    pages = "151",
    year = "2021"
}

@article{Yao:2018qjv,
    author = "Yao, J. M. and Engel, J. and Wang, L. J. and Jiao, C. F. and Hergert, H.",
    title = "{Generator-coordinate reference states for spectra and $0\nu\beta\beta$ decay in the in-medium similarity renormalization group}",
    eprint = "1807.11053",
    archivePrefix = "arXiv",
    primaryClass = "nucl-th",
    doi = "10.1103/PhysRevC.98.054311",
    journal = "Phys. Rev. C",
    volume = "98",
    number = "5",
    pages = "054311",
    year = "2018"
}

@article{Yao:2019rck,
    author = "Yao, J. M. and Bally, B. and Engel, J. and Wirth, R. and Rodr{\'\i}guez, T. R. and Hergert, H.",
    title = "{$Ab Initio$ Treatment of Collective Correlations and the Neutrinoless Double Beta Decay of $^{48}$Ca}",
    eprint = "1908.05424",
    archivePrefix = "arXiv",
    primaryClass = "nucl-th",
    doi = "10.1103/PhysRevLett.124.232501",
    journal = "Phys. Rev. Lett.",
    volume = "124",
    number = "23",
    pages = "232501",
    year = "2020"
}

@article{Frosini:2021fjf,
    author = "Frosini, Mikael and Duguet, Thomas and Ebran, Jean-Paul and Som{\`a}, Vittorio",
    title = "{Multi-reference many-body perturbation theory for nuclei: I. Novel PGCM-PT formalism}",
    eprint = "2110.15737",
    archivePrefix = "arXiv",
    primaryClass = "nucl-th",
    doi = "10.1140/epja/s10050-022-00692-z",
    journal = "Eur. Phys. J. A",
    volume = "58",
    number = "4",
    pages = "62",
    year = "2022"
}

@article{Frosini:2021sxj,
    author = "Frosini, Mikael and Duguet, Thomas and Ebran, Jean-Paul and Bally, Benjamin and Mongelli, Tobias and Rodr{\'\i}guez, Tom{\'a}s R. and Roth, Robert and Som{\`a}, Vittorio",
    title = "{Multi-reference many-body perturbation theory for nuclei: II. Ab initio study of neon isotopes via PGCM and IM-NCSM calculations}",
    eprint = "2111.00797",
    archivePrefix = "arXiv",
    primaryClass = "nucl-th",
    doi = "10.1140/epja/s10050-022-00693-y",
    journal = "Eur. Phys. J. A",
    volume = "58",
    number = "4",
    pages = "63",
    year = "2022"
}

@article{Frosini:2021ddm,
    author = "Frosini, Mikael and Duguet, Thomas and Ebran, Jean-Paul and Bally, Benjamin and Hergert, Heiko and Rodr{\'\i}guez, Tom{\'a}s R. and Roth, Robert and Yao, Jiangming and Som{\`a}, Vittorio",
    title = "{Multi-reference many-body perturbation theory for nuclei: III. Ab initio calculations at second order in PGCM-PT}",
    eprint = "2111.01461",
    archivePrefix = "arXiv",
    primaryClass = "nucl-th",
    doi = "10.1140/epja/s10050-022-00694-x",
    journal = "Eur. Phys. J. A",
    volume = "58",
    number = "4",
    pages = "64",
    year = "2022"
}

@article{Nijs:2020ors,
    author = {Nijs, Govert and van der Schee, Wilke and G{\"u}rsoy, Umut and Snellings, Raimond},
    title = "{Transverse Momentum Differential Global Analysis of Heavy-Ion Collisions}",
    eprint = "2010.15130",
    archivePrefix = "arXiv",
    primaryClass = "nucl-th",
    reportNumber = "CERN-TH-2020-174, MIT-CTP/5250",
    doi = "10.1103/PhysRevLett.126.202301",
    journal = "Phys. Rev. Lett.",
    volume = "126",
    number = "20",
    pages = "202301",
    year = "2021"
}

@article{Nijs:2020roc,
    author = {Nijs, Govert and van der Schee, Wilke and G{\"u}rsoy, Umut and Snellings, Raimond},
    title = "{Bayesian analysis of heavy ion collisions with the heavy ion computational framework Trajectum}",
    eprint = "2010.15134",
    archivePrefix = "arXiv",
    primaryClass = "nucl-th",
    reportNumber = "CERN-TH-2020-175, MIT-CTP/5251",
    doi = "10.1103/PhysRevC.103.054909",
    journal = "Phys. Rev. C",
    volume = "103",
    number = "5",
    pages = "054909",
    year = "2021"
}

@article{Huss:2020whe,
    author = "Huss, Alexander and Kurkela, Aleksi and Mazeliauskas, Aleksas and Paatelainen, Risto and van der Schee, Wilke and Wiedemann, Urs Achim",
    title = "{Predicting parton energy loss in small collision systems}",
    eprint = "2007.13758",
    archivePrefix = "arXiv",
    primaryClass = "hep-ph",
    reportNumber = "CERN-TH-2020-127",
    doi = "10.1103/PhysRevC.103.054903",
    journal = "Phys. Rev. C",
    volume = "103",
    number = "5",
    pages = "054903",
    year = "2021"
}

@article{Arnold:2008iy,
    author = "Arnold, Peter Brockway",
    title = "{Simple Formula for High-Energy Gluon Bremsstrahlung in a Finite, Expanding Medium}",
    eprint = "0808.2767",
    archivePrefix = "arXiv",
    primaryClass = "hep-ph",
    doi = "10.1103/PhysRevD.79.065025",
    journal = "Phys. Rev. D",
    volume = "79",
    pages = "065025",
    year = "2009"
}

@article{Baier:1996kr,
    author = "Baier, R. and Dokshitzer, Yuri L. and Mueller, Alfred H. and Peigne, S. and Schiff, D.",
    title = "{Radiative energy loss of high-energy quarks and gluons in a finite volume quark - gluon plasma}",
    eprint = "hep-ph/9607355",
    archivePrefix = "arXiv",
    reportNumber = "BI-TP-96-21, CUTP-759, LPTHE-ORSAY-96-34",
    doi = "10.1016/S0550-3213(96)00553-6",
    journal = "Nucl. Phys. B",
    volume = "483",
    pages = "291--320",
    year = "1997"
}

@article{Zakharov:1996fv,
    author = "Zakharov, B. G.",
    title = "{Fully quantum treatment of the Landau-Pomeranchuk-Migdal effect in QED and QCD}",
    eprint = "hep-ph/9607440",
    archivePrefix = "arXiv",
    doi = "10.1134/1.567126",
    journal = "JETP Lett.",
    volume = "63",
    pages = "952--957",
    year = "1996"
}

@article{Eskola:2016oht,
    author = "Eskola, Kari J. and Paakkinen, Petja and Paukkunen, Hannu and Salgado, Carlos A.",
    title = "{EPPS16: Nuclear parton distributions with LHC data}",
    eprint = "1612.05741",
    archivePrefix = "arXiv",
    primaryClass = "hep-ph",
    doi = "10.1140/epjc/s10052-017-4725-9",
    journal = "Eur. Phys. J. C",
    volume = "77",
    number = "3",
    pages = "163",
    year = "2017"
}

@article{Moreland:2014oya,
    author = "Moreland, J. Scott and Bernhard, Jonah E. and Bass, Steffen A.",
    title = "{Alternative ansatz to wounded nucleon and binary collision scaling in high-energy nuclear collisions}",
    eprint = "1412.4708",
    archivePrefix = "arXiv",
    primaryClass = "nucl-th",
    doi = "10.1103/PhysRevC.92.011901",
    journal = "Phys. Rev. C",
    volume = "92",
    number = "1",
    pages = "011901",
    year = "2015"
}

@article{Beattie:2022ojg,
    author = "Beattie, Caitlin and Nijs, Govert and Sas, Mike and van der Schee, Wilke",
    title = "{Hard probe path lengths and event-shape engineering of the quark-gluon plasma}",
    eprint = "2203.13265",
    archivePrefix = "arXiv",
    primaryClass = "nucl-th",
    reportNumber = "CERN-TH-2022-051, MIT-CTP/5397",
    doi = "10.1016/j.physletb.2022.137596",
    journal = "Phys. Lett. B",
    volume = "836",
    pages = "137596",
    year = "2023"
}

@article{Kolbe:2023rsq,
    author = "Kolb{\'e}, Isobel",
    title = "{JEWEL on a (2+1)D background with applications to small systems and substructure}",
    eprint = "2303.14166",
    archivePrefix = "arXiv",
    primaryClass = "nucl-th",
    month = "3",
    year = "2023"
}

@article{Zapp:2012ak,
    author = "Zapp, Korinna C. and Krauss, Frank and Wiedemann, Urs A.",
    title = "{A perturbative framework for jet quenching}",
    eprint = "1212.1599",
    archivePrefix = "arXiv",
    primaryClass = "hep-ph",
    reportNumber = "CERN-PH-TH-2012-344, IPPP-12-92, DCPT-12-84, MCNET-12-14",
    doi = "10.1007/JHEP03(2013)080",
    journal = "JHEP",
    volume = "03",
    pages = "080",
    year = "2013"
}

@article{Zapp:2013vla,
    author = "Zapp, Korinna C.",
    title = "{JEWEL 2.0.0: directions for use}",
    eprint = "1311.0048",
    archivePrefix = "arXiv",
    primaryClass = "hep-ph",
    reportNumber = "CERN-PH-TH-2013-255, MCNET-13-17",
    doi = "10.1140/epjc/s10052-014-2762-1",
    journal = "Eur. Phys. J. C",
    volume = "74",
    number = "2",
    pages = "2762",
    year = "2014"
}

@article{Zapp:2008gi,
    author = "Zapp, Korinna and Ingelman, Gunnar and Rathsman, Johan and Stachel, Johanna and Wiedemann, Urs Achim",
    title = "{A Monte Carlo Model for 'Jet Quenching'}",
    eprint = "0804.3568",
    archivePrefix = "arXiv",
    primaryClass = "hep-ph",
    reportNumber = "CERN-PH-TH-2008-067",
    doi = "10.1140/epjc/s10052-009-0941-2",
    journal = "Eur. Phys. J. C",
    volume = "60",
    pages = "617--632",
    year = "2009"
}

@article{Eskola:2021nhw,
    author = "Eskola, Kari J. and Paakkinen, Petja and Paukkunen, Hannu and Salgado, Carlos A.",
    title = "{EPPS21: a global QCD analysis of nuclear PDFs}",
    eprint = "2112.12462",
    archivePrefix = "arXiv",
    primaryClass = "hep-ph",
    doi = "10.1140/epjc/s10052-022-10359-0",
    journal = "Eur. Phys. J. C",
    volume = "82",
    number = "5",
    pages = "413",
    year = "2022"
}

@article{PHENIX:2015siv,
    author = "Adare, A. and others",
    collaboration = "PHENIX",
    title = "{An Upgrade Proposal from the PHENIX Collaboration}",
    eprint = "1501.06197",
    archivePrefix = "arXiv",
    primaryClass = "nucl-ex",
    month = "1",
    year = "2015"
}

@article{ATLAS:2018gwx,
    author = "Aaboud, Morad and others",
    collaboration = "ATLAS",
    title = "{Measurement of the nuclear modification factor for inclusive jets in Pb+Pb collisions at $\sqrt{s_\mathrm{NN}}=5.02$ TeV with the ATLAS detector}",
    eprint = "1805.05635",
    archivePrefix = "arXiv",
    primaryClass = "nucl-ex",
    reportNumber = "CERN-EP-2018-105",
    doi = "10.1016/j.physletb.2018.10.076",
    journal = "Phys. Lett. B",
    volume = "790",
    pages = "108--128",
    year = "2019"
}

@article{ALICE:2023waz,
    author = "Acharya, Shreyasi and others",
    collaboration = "ALICE",
    title = "{Measurement of the radius dependence of charged-particle jet suppression in Pb{\textendash}Pb collisions at sNN=5.02TeV}",
    eprint = "2303.00592",
    archivePrefix = "arXiv",
    primaryClass = "nucl-ex",
    reportNumber = "CERN-EP-2023-027",
    doi = "10.1016/j.physletb.2023.138412",
    journal = "Phys. Lett. B",
    volume = "849",
    pages = "138412",
    year = "2024"
}

@article{Caron-Huot:2010qjx,
    author = "Caron-Huot, Simon and Gale, Charles",
    title = "{Finite-size effects on the radiative energy loss of a fast parton in hot and dense strongly interacting matter}",
    eprint = "1006.2379",
    archivePrefix = "arXiv",
    primaryClass = "hep-ph",
    doi = "10.1103/PhysRevC.82.064902",
    journal = "Phys. Rev. C",
    volume = "82",
    pages = "064902",
    year = "2010"
}

@article{Mazeliauskas:2025clt,
    author = "Mazeliauskas, Aleksas",
    title = "{Energy loss baseline for light hadrons in oxygen-oxygen collisions at $\sqrt{s_\mathrm{NN}}=5.36\,\text{TeV}$}",
    eprint = "2509.07008",
    archivePrefix = "arXiv",
    primaryClass = "hep-ph",
    month = "9",
    year = "2025"
}

@article{Andres:2022bql,
    author = "Andres, Carlota and Apolin{\'a}rio, Liliana and Dominguez, Fabio and Martinez, Marcos Gonzalez and Salgado, Carlos A.",
    title = "{Medium-induced radiation with vacuum propagation in the pre-hydrodynamics phase}",
    eprint = "2211.10161",
    archivePrefix = "arXiv",
    primaryClass = "hep-ph",
    doi = "10.1007/JHEP03(2023)189",
    journal = "JHEP",
    volume = "03",
    pages = "189",
    year = "2023"
}

@article{Jonas:2026yoz,
    author = "Jonas, Florian and Loizides, Constantin and Mazeliauskas, Aleksas and Paakkinen, Petja and Strangmann, Nicolas",
    title = "{A compendium of cold-nuclear matter baseline predictions in light-ion collisions}",
    eprint = "2602.15928",
    archivePrefix = "arXiv",
    primaryClass = "hep-ph",
    month = "2",
    year = "2026"
}

@article{Milhano:2015mng,
    author = "Milhano, Jos{\'e} Guilherme and Zapp, Korinna Christine",
    title = "{Origins of the di-jet asymmetry in heavy ion collisions}",
    eprint = "1512.08107",
    archivePrefix = "arXiv",
    primaryClass = "hep-ph",
    reportNumber = "CERN-PH-TH-2015-310, MCNET-15-31",
    doi = "10.1140/epjc/s10052-016-4130-9",
    journal = "Eur. Phys. J. C",
    volume = "76",
    number = "5",
    pages = "288",
    year = "2016"
}

@article{CMS:2025bta,
    author = "Hayrapetyan, Aram and others",
    collaboration = "CMS",
    title = "{Observation of Suppressed Charged-Particle Production in Ultrarelativistic Oxygen-Oxygen Collisions}",
    eprint = "2510.09864",
    archivePrefix = "arXiv",
    primaryClass = "nucl-ex",
    reportNumber = "CMS-HIN-25-008, CERN-EP-2025-226",
    doi = "10.1103/89sf-9t1x",
    journal = "Phys. Rev. Lett.",
    volume = "136",
    number = "16",
    pages = "162301",
    year = "2026"
}

@dataset{van_der_schee_2025_22257566,
  author       = {van der Schee, Wilke and
                  Kolbé, Isobel and
                  Nijs, Govert and
                  Iqbal, Shahin},
  title        = {Three Predictions for Partonic Energy Loss from
                   Light to Heavy Ion Collisions
                  },
  month        = sep,
  year         = 2025,
  publisher    = {Zenodo},
  doi          = {10.5281/zenodo.22257566},
  url          = {https://doi.org/10.5281/zenodo.22257566},
}

\appendix
\section{Systematic Effects}

    In this section, we briefly discuss the different systematic effects taken into account in each model, summarised in \cref{tab:systematicEffects}.
    % By ``coherence'' we mean the effect of coherent radiation, which is really only possible in JEWEL.

    Energy loss is known to be sensitive to its starting time \cite{Andres:2022bql}.
    In \jewel{} vacuum radiation before the initialisation time of the hydrodynamic background is included, while medium-induced energy-loss is started only in the hydrodynamic phase.
    In the Pathlength model, the energy loss is computed as a temperature-dependent quantity, so here the energy loss is naturally also started in the hydrodynamic phase.

    \begin{table}
        \centering
        \begin{tabular}{|p{1.6cm}|p{1.5cm}|p{1.5cm}|p{1.5cm}|p{1.5cm}|}
            \hline
             Model      & E-loss model & Medium     & nPDFs     & Energy-loss start time [fm] \\
            \hline
            \hline
             Simple     & Rad.    & Average    & EPPS16    &  0.05 \\
             \hline 
             Pathlength & Pheno.    & EbyE dyn.    &     ---     & $0.40 \pm 0.17$\\
             \hline
             JEWEL      & Rad. \& elas.    & EbyE dyn.    & EPPS21   & 0.56 \\
             \hline
                     \end{tabular}
        \caption{Summary of the systematic effects taken into account in each model.}
        \label{tab:systematicEffects}
    \end{table}

\begin{figure}[t]
        \centering
        \includegraphics[width=8.5cm]{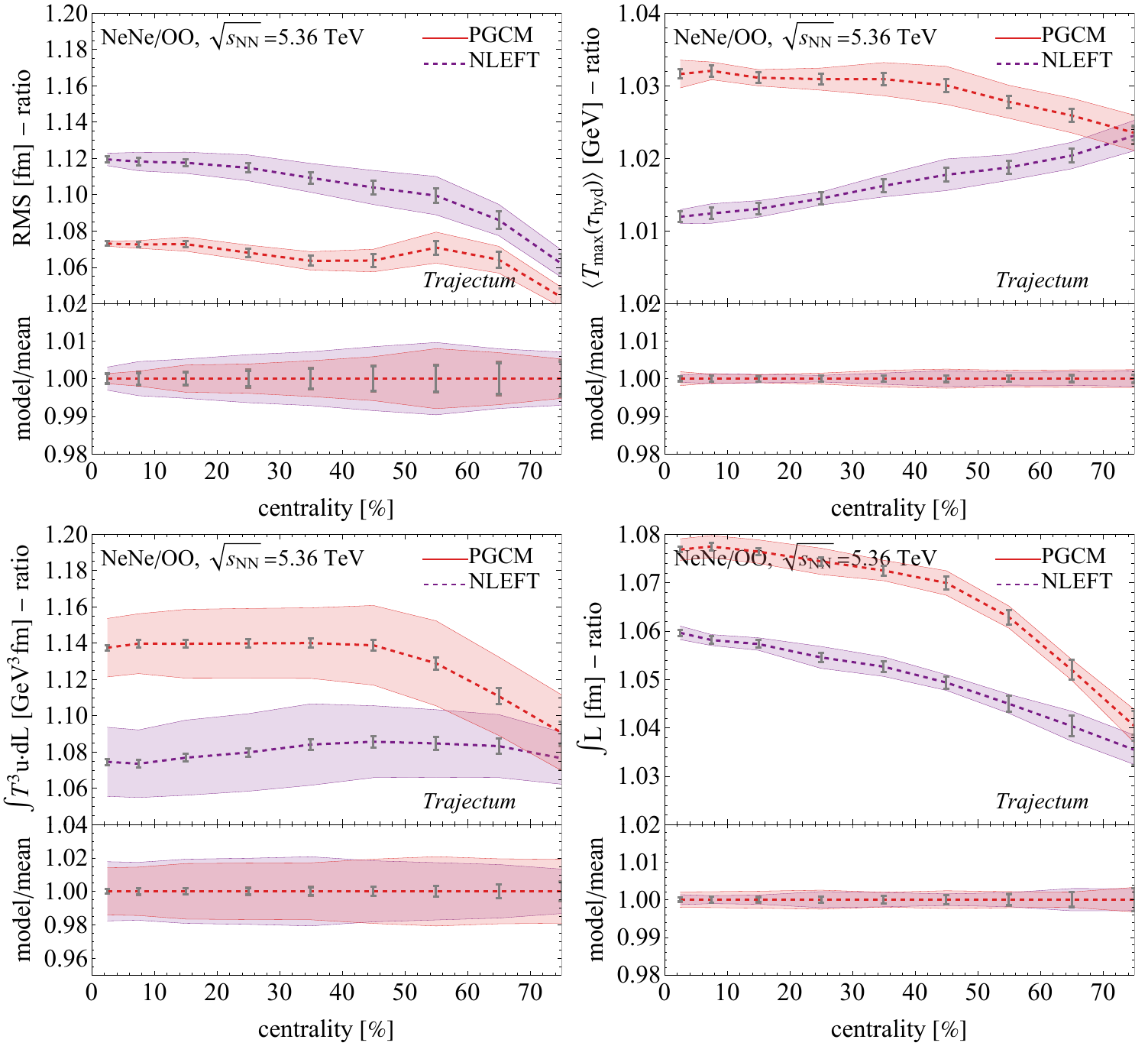}\\
        \caption{We show the \emph{Trajectum} \nene{} over \oooo{} ratios for respectively the initial size of the QGP, the maximum temperature, the temperature-weighted pathlength as well as the pathlength itself.
        \label{fig:trajectumratios}}
    \end{figure}

In Fig.~\ref{fig:trajectumratios} we give extra information on the \nene{} over \oooo{} ratios regarding the pathlengths for both PGCM and NLEFT configurations used in the main text. Already in \cite{Giacalone:2024luz} it was noticed that the difference in size (measured by the charge radius) is bigger for the NLEFT configurations (\nee{} is 1.14 times bigger than \oo{}) than for the PGCM configurations (a factor 1.08). Since the experimental value is 1.115 we expect the true ratio of the sizes to be somewhere between the NLEFT and PGCM configurations.

This size difference is apparent in the RMS of the initial QGP droplet, especially for central collisions (Fig.~\ref{fig:trajectumratios} top-left).  After the collision the \nee{} QGP droplet is naturally hotter since it has more nucleons. Nevertheless, the smaller size difference in PGCM means this temperature difference is more pronounced in PGCM (Fig.~\ref{fig:trajectumratios} top-right).

For the (temperature-weighted) paths this has the important consequence that the paths are longer for PGCM (Fig.~\ref{fig:trajectumratios} (bottom-left) and bottom-right for (temperature-weighted) paths). We note that even without temperature-weighting the pathlenghts in PGCM are longer since the plasma lives longer and not because the plasma started larger.

\begin{figure}
\includegraphics[width=8cm]{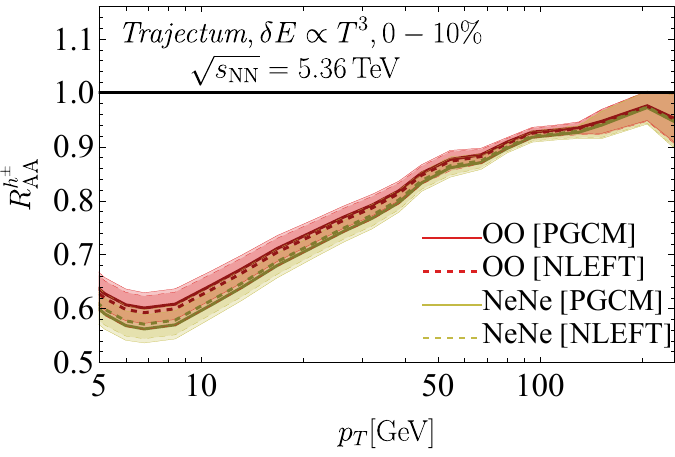}
\caption{Equivalent figure of \cref{fig:RAAMB} (top), but
now in the 0-10\% centrality class.\label{fig:RAA010}}
\end{figure}

Lastly, in Fig.~\ref{fig:RAA010} we present the centrality selected nuclear modification factor for the Pathlength model.

\section{Jets with JEWEL}

    \begin{figure}
        \centering
        \includegraphics[width=8cm]{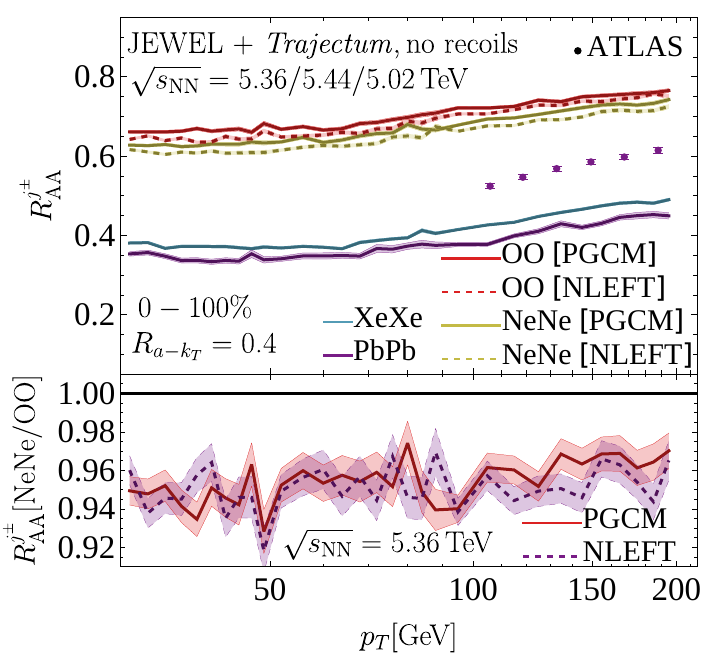}
        \caption{Nuclear modification factor for $R=0.4$ jets with JEWEL and \emph{Trajectum} for minimum bias events. We include the ATLAS data for PbPb collisions \cite{ATLAS:2018gwx}.}
        \label{fig:jewelraajetsMB04}
    \end{figure}

    \begin{figure}
        \centering
        \includegraphics[width=8cm]{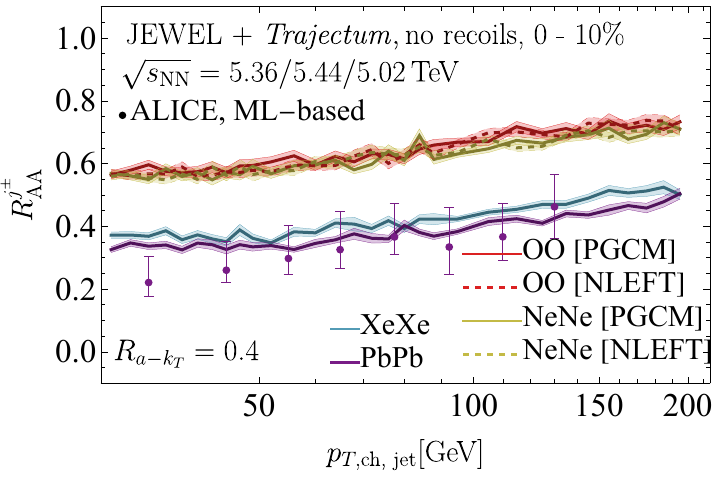}
        \caption{Nuclear modification factor for $R=0.4$ charged jets with JEWEL and \emph{Trajectum} for 0 - 10\% central events. We include the ALICE data for PbPb collisions \cite{ALICE:2023waz}.}
        \label{fig:jewelraajets04}
    \end{figure}

In the main text, we noted that JEWEL is primarily designed for jet observables rather than inclusive charged hadron spectra. 
\Cref{fig:jewelraajetsMB04} shows the $R = 0.4$ charged particle anti-$k_T$ jet nuclear modification factor computed with the JEWEL-\textit{Trajectum} interface, compared with ATLAS data for PbPb collisions \cite{ATLAS:2018gwx}. In this figure we averaged the ATLAS centrality classes weighted by the number of binary collisions and included the missing 80-100\% as a (tiny) uncertainty.
While JEWEL captures some qualitative features of jet quenching, quantitative agreement with the ATLAS jet data remains challenging with the default parameter settings used throughout this work. 
The discrepancy reflects known tensions \cite{PHENIX:2015siv,Zapp:2012ak} in simultaneously describing both the magnitude of jet suppression and the inclusive hadron spectrum within JEWEL's current framework, particularly the kinematic effects from elastic scatterings discussed in \cref{sec:jeweltrajectum}.
Given these tensions in the baseline PbPb system, we refrain from fine-tuning JEWEL's parameters, as this would introduce additional model dependence without clear physical justification. 
We also note that comparing to charged jets for central PbPb collisions compared with ALICE \cite{ALICE:2023waz} does agree well with the JEWEL-\textit{Trajectum} results as shown in Fig.~\ref{fig:jewelraajets04}.
Nevertheless, the ion-size dependence and the qualitative behaviour of the NeNe/OO ratio shown in the main text remain robust features of the model, reflecting the underlying path-length dependence of energy loss mechanisms implemented in JEWEL.
Future developments of the JEWEL-\textit{Trajectum} interface, potentially including additional tuning and the inclusion of recoils, may improve the simultaneous description of both jet and hadron observables across different collision systems. 
In addition to such tunings, since the first submission of the present manuscript, a new version, JEWEL-2.6.0, has become public, in which more tunable physics in JEWEL's handling of coherence effects have been implemented, offering a particularly interesting avenue for future studies.

\end{document}